\documentclass[aps,prb,reprint,amsmath,amssymb,superscriptaddress]{revtex4-2}
\usepackage[bookmarks=false,linkcolor=blue,urlcolor=blue,colorlinks,citecolor=blue]{hyperref}
\usepackage{graphicx}
\usepackage{amsmath}
\usepackage[svgnames]{xcolor}
\usepackage{dcolumn}
\usepackage{bm}
\usepackage{scalerel}
\usepackage{textcomp}
\usepackage{physics}
\usepackage{dsfont}
\usepackage{booktabs}

\newcommand{\figref}[2]{\hyperref[#1]{\ref{#1}(#2)}}
\newcommand{\figsref}[2]{\hyperref[#1]{\ref{#1}#2}}
\newcommand{\Weizmann}{Department of Condensed Matter Physics, Weizmann Institute of Science, Rehovot, Israel 7610001}
\newcommand{\McGill}{Department of Physics, McGill University, Montr\'{e}al, Qu\'{e}bec H3A 2T8, Canada}
\newcommand{\EqualContribution}{G.~M. and O.~L. contributed equally to the work.}

\begin{document}

\title{RG-Inspired Neural Networks for Computing Topological Invariants}

\author{Gilad Margalit}
\thanks{\EqualContribution}
\affiliation{\Weizmann}

\author{Omri Lesser}
\thanks{\EqualContribution}
\affiliation{\Weizmann}

\author{T. Pereg-Barnea}
\affiliation{\McGill}

\author{Yuval Oreg}
\affiliation{\Weizmann}

\begin{abstract}
We show that artificial neural networks (ANNs) can, to high accuracy, determine the topological invariant of a disordered system given its two-dimensional real-space Hamiltonian. Furthermore, we describe a ``renormalization-group" (RG) network, an ANN which converts a Hamiltonian on a large lattice to another on a small lattice while preserving the invariant. By iteratively applying the RG network to a ``base" network that computes the Chern number of a small lattice of set size, we are able to process larger lattices without re-training the system. We therefore show that it is possible to compute real-space topological invariants for systems larger than those on which the network was trained. 
This opens the door for computation times significantly faster and more scalable than previous methods.
\end{abstract}
\maketitle

\begin{figure*}[t]
    \centering
    \includegraphics[width=\textwidth]{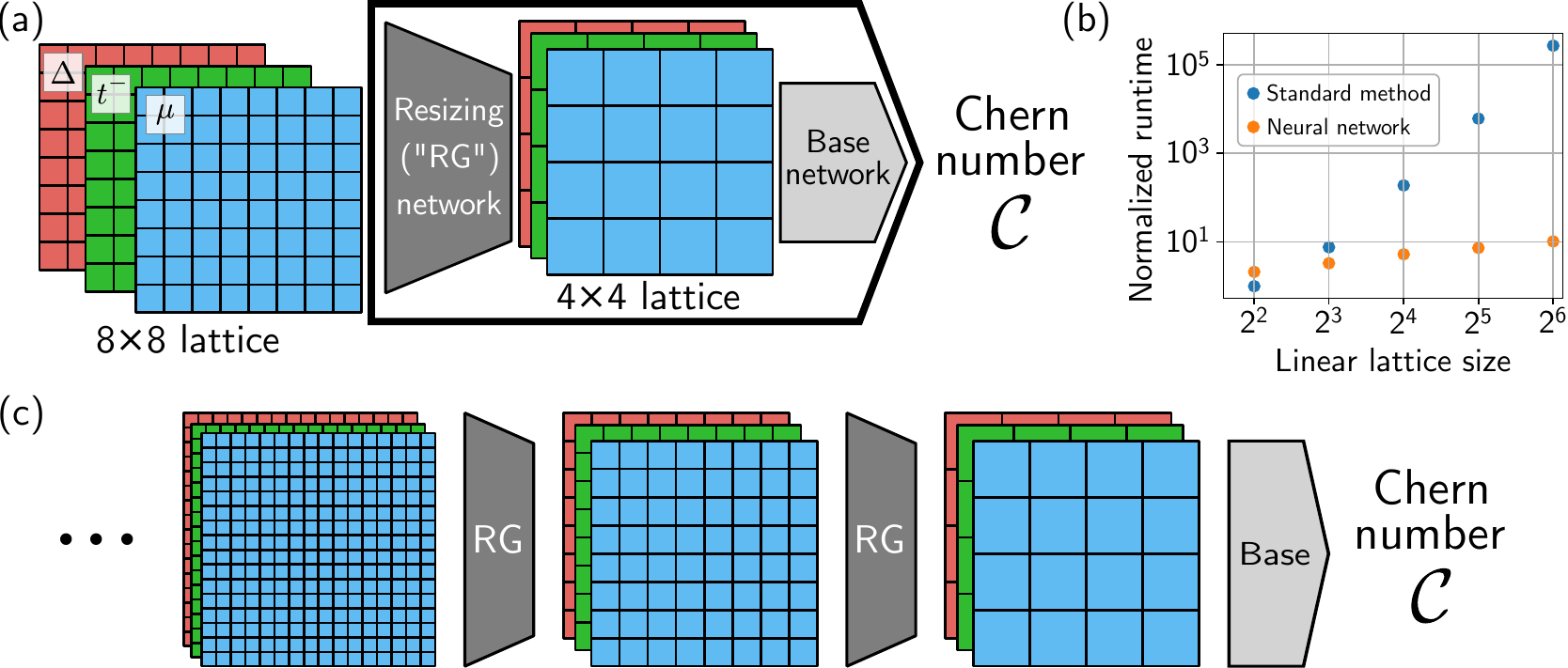}
    \caption{(a)~RG and base network block structure. In the first stage, the base network is trained to find the Chern numbers of $4 \times 4$ lattices. Then a compound network (thick outline), which includes the pre-trained base network and a resizing ``RG" network, is trained on $8 \times 8$ lattices. The RG network reduces the lattice dimensions by 2 while preserving the Chern number, allowing it to be solved by the base network. (b)~Runtime scaling comparison between the Bott index method and successive application of an RG network. The times for both methods are normalized by the runtime for the $4 \times 4$ case calculated using the Bott index method. The Bott index method, used as a baseline in this work, is $\mathcal{O}(n^4)$ to $\mathcal{O}(n^6)$, depending on the efficiency of diagonalization methods, where $n$ is the linear lattice size. Extreme parallelization allows the RG method to run in sub-linear time. Notice that the accuracy of the result is not shown; it is likely that the accuracy of our method declines for particularly large lattices, though additional research may change this. (c)~Extension of the diagram in (a) to arbitrary lattice sizes by iteratively applying the RG network, then computing the Chern number via the base network once the lattice has been reduced to size $4 \times 4$.
    \label{fig:block_diagram}}
\end{figure*}

\section{Introduction}
Machine learning (ML) has been redefining the way computer-related problems are solved~\cite{goodfellow_deep_2016,nielsen_neural_2015}.
Until about a decade ago, the methodology of algorithmic problem solving had been quite clear: a human programmer looks for rules and regularities, which are then transformed into conditional statements run by the computer.
Machine learning takes a different approach, based on the concept of \emph{training}, i.e., repeatedly changing the values of internal parameters in the algorithm in order to reach a desired goal. 
After the training, the ML algorithm finds a solution by itself, earning it the name \emph{artificial intelligence}.
Owing to massive improvements in both software and hardware, machine learning has emerged as a game-changing approach to formidably difficult problems, well beyond the reach of traditional algorithms.
Popular examples of such problems are image and pattern recognition~\cite{pak_review_2017} and natural language processing~\cite{nadkarni_natural_2011,mankolli_machine_2020}.

The great promise held by machine learning did not evade the notice of physicists~\cite{mehta_high-bias_2018,carleo_machine_2019,schmidt_recent_2019,udrescu_ai_2020}. 
ML approaches are now widespread in many areas of experimental physics involving the analysis of large amounts of data~\cite{radovic_machine_2018,guest_deep_2018,bourilkov_machine_2019-1,lustig_identifying_2020,kaming_unsupervised_2021}. 
Additionally, ML techniques---and in particular, artificial neural networks (ANNs)---keep finding applications in theoretical condensed matter physics~\cite{carrasquilla_machine_2020}. 
For example, ANNs have been used as a means to detect phase transitions~\cite{venderley_machine_2018,sun_deep_2018,beach_machine_2018,van_nieuwenburg_learning_2018,kuo_unsupervised_2021} and solve quantum many-body problems~\cite{carleo_solving_2017,chng_machine_2017,broecker_machine_2017,koch-janusz_mutual_2018,li_neural_2018,luo_backflow_2019}. 
Typically, a representation of the system is chosen (like Hamiltonian parameters or wavefunction), and the ML algorithm learns to calculate a desired property (like phase classification, ground-state energy, or average magnetization).
An important advantage of ML techniques in solving such problems is the sizeable speedups they offer compared to traditional methods.

With the advent of topological phases of matter in the last several decades~\cite{volovik_universe_2009,hasan_colloquium_2010, moore_birth_2010,qi_topological_2011,bernevig_topological_2013}, calculating topological invariants has become a highly relevant problem in contemporary condensed matter physics~\cite{thouless_quantized_1982,niu_quantized_1985,schnyder_classification_2008,hatsugai_chern_1993,fukui_chern_2005,fulga_scattering_2012}.
These integer-valued invariants are important markers of the bulk topology and the edge properties. 
A famous example is the quantum Hall effect~\cite{klitzing_new_1980,tsui_two-dimensional_1982}, where the relevant topological invariant---the Chern number---counts the number of chiral edge modes~\cite{thouless_quantized_1982,girvin_quantum_1999}.
In translation-invariant and non-interacting systems, calculating topological invariants is usually a straightforward task~\cite{berry_quantal_1984,bernevig_topological_2013}.
However, the presence of disorder alters the picture and makes the calculation challenging.
In two spatial dimensions, there are several methods for solving this problem~\cite{loring_disordered_2011,borchmann_anderson_2016}, but their scaling with the system size makes them unfeasible for large systems.
It is therefore tempting to utilize ML techniques to tackle this problem.
In particular, since topological invariants are integers, the task at hand is actually a classification problem, for which ML algorithms are particularly well-suited~\cite{ohtsuki_deep_2016,deng_machine_2017,zhang_machine_2018,carvalho_real-space_2018,yoshioka_learning_2018,rodriguez-nieva_identifying_2019,holanda_machine_2020,scheurer_unsupervised_2020,mertz_engineering_2021}. 

Despite the apparent appeal and compatibility of ML techniques to the task of calculating topological invariants of large disordered systems, there is one clear drawback. The overwhelming majority of ML algorithms assume a constant input size, whereas we would like our method to be general---we do not wish to train a separate ANN for each system size.
Furthermore, generating labeled samples and training an ANN both become computationally intractable, for large lattice sizes.

To overcome these hurdles, we employ a two-stage solution strategy, as illustrated in Fig.~\figref{fig:block_diagram}{a}. First, we train an ANN to solve the simple problem of calculating the Chern number of a small system with a fixed size. We shall refer to this as the ``base" network.  Then, we train a second ANN whose goal is to map a large system to a smaller one, while preserving the Chern number. This resizing network applies a non-linear transformation to the original system and reduces its linear size by a factor of two. Crucially the network is capable of reducing any $N \times N$ system to an effective $\frac{N}{2}\times\frac{N}{2}$ one. It then becomes possible to start with a large system and iteratively apply the resizing network, without further training, until it reaches the fixed small size accepted by the base network. At this point, the reduced system is fed into the base network, which outputs its Chern number. The iterative process is demonstrated in Fig.~\figref{fig:block_diagram}{c}. This architecture allows for a very substantial speedup in computation time compared to the standard methods, as shown in Fig.~\figref{fig:block_diagram}{b}. Though the base network is not the source of the computational speedup, expressing the base Chern solver as a neural network allows easier integration with the RG network during training.

Mapping a large system onto a smaller effective one is a well-established paradigm in condensed matter physics, and the overarching theoretical framework for doing so is the renormalization group (RG)~\cite{hu_introduction_1982,white_real-space_1992,refael_strong_2013}. In particular, our approach is similar in spirit to real-space RG or block-spin decimation~\cite{kadanoff_scaling_1966,wilson_renormalization_1971}. There, the idea is to cluster neighboring degrees of freedom, and treat them as the effective degrees of freedom in a new, coarse-grained system. Such methods have had some success in elementary problems, but they are typically insufficient for capturing more complex situations. They may fail by generating new types of interactions that were absent in the original system, or by ignoring the entanglement between coarse grained blocks---the ground state of a large system is not necessarily composed of the ground states of each block. Here, we aim to address the difficulties common in real-space RG methods by assuming very little about the resizing transformation (which we refer to as the ``RG" network): the high versatility and non-linearity of the ANN allow it to realize extremely complicated transformations. The network, by construction, \emph{learns} an appropriate RG-like transformation that preserves the topological invariant. Furthermore, the built-in separation into network blocks allows us to gain some insight into the nature of the learned transformation. Such insight is not typically possible in ANNs (see Appendix~\ref{interpretability}).

\begin{figure*}[t!]
    \centering
    \includegraphics[width=\textwidth]{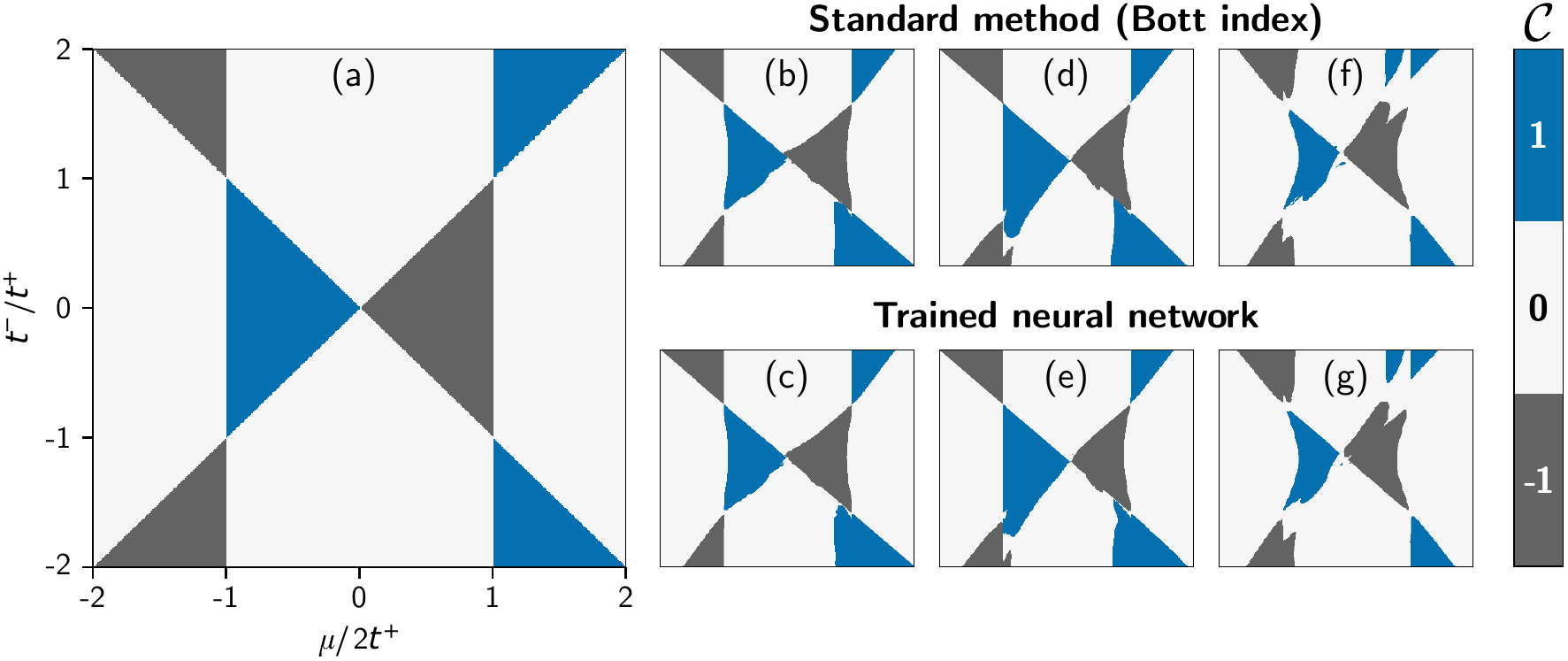}
    \caption{(a)~Phase diagram of a clean system (every site identical) in terms of the chemical potential $\mu$ and hopping anisotropy $t^-$. Color indicates the Chern number ${\cal C}$ of the system. (b), (d), (f)~Examples of disordered phase diagrams of $4\times 4$ lattices, calculated using the Bott index method~\cite{loring_disordered_2011}. (c), (e), (g) are the corresponding phase diagrams calculated by the base neural network
    \label{fig:phase_diagrams}}
\end{figure*}

\section{System and base network}
Our example system is a disordered $p + ip$ superconductor in real space~\cite{alicea_new_2012,bernevig_topological_2013,mross_theory_2018}, described by the following tight-binding Hamiltonian:
\begin{equation}
\begin{aligned}
\label{eq:Hamiltonian}
H = \sum_{i,j} \Big[& \mu_{i,j}c^\dagger_{i,j}c_{i,j} + t^x_{i,j}c^\dagger_{i+1,j}c_{i,j} + t^y_{i,j}c^\dagger_{i,j+1}c_{i,j} \\
&+ \Delta_{i,j}c^\dagger_{i+1,j}c^\dagger_{i,j} +
i\Delta_{i,j}c^\dagger_{i,j+1}c^\dagger_{i,j} + \text{H.c.} \Big],
\end{aligned}
\end{equation}
where $c^\dagger_{i,j}$,  $c_{i,j}$ are creation and annihilation operators for an electron on the site $(i,j)$ of the square lattice with periodic boundary conditions. Each site has an on-site energy $\mu$, hopping energies $t^x$ and $t^y$ to adjacent sites, and a superconducting pair potential $\Delta$ with $p + ip$ symmetry.
It is helpful to define $t^{\pm}_{i,j} \equiv \left(t^x_{i,j} \pm t^y_{i,j}\right)/2$, and normalize the parameters such that $t^+=1$ for all sites. This leaves us with three parameters per site: $\mu_{i,j}$, $t^-_{i,j}$, and $\Delta_{i,j}$ (the latter two are bond parameters, but they can equivalently be regarded as site parameters).

In the clean case the Chern number is independent of $\Delta$, and it is determined by the uniform values of $\mu$ and $t^-$~\cite{mross_theory_2018}, as shown in the phase diagram in Fig.~\figref{fig:phase_diagrams}{a}.
In the disordered case, we can construct a similar phase diagram.
A disorder realization is defined by a list of random values of $\mu$, $t^{-}$, and $\Delta$ at each site with zero mean. 
On top of that, we uniformly vary the average values of $\mu$ and $t^{-}$, and calculate the real-space Chern number for the resulting lattice via the Bott index method~\cite{loring_disordered_2011}.
Figures.~\figsref{fig:phase_diagrams}{(b), (d), (f)} each correspond to a single disorder realization for a $4 \times 4$ lattice.

We shall now reproduce this calculation using an ANN.
Our base neural network acts on a $4 \times 4$ lattice, and thus takes as input a set of 48 numbers: 16 sites times 3 ``channels" per site: $\mu$, $t^{-}$, $\Delta$.
We used a convolutional ResNet architecture with 9 layers, which proved fast to train (see Appendix~\ref{sec:base} for more details on the network structure, loss function, and training process).
The network was trained on $10^8$ disorder realizations, whose Chern number was calculated using the Bott index method and treated as a label.
For each sample, the network was trained to minimize the difference between its estimate of the Chern number and the sample's label. This process continued until the network reached a maximum accuracy of 99.02\%.

As a visualization of the network's performance, we computed the phase diagrams corresponding to several disorder realizations.
This was done by feeding the disorder configuration with the uniform $\mu$, $t^{-}$ offsets into the trained network.
The resulting phase diagrams, shown in Figs.~\figsref{fig:phase_diagrams}{(c), (e), (g)}, are in very good agreement with the directly computed ones shown in Fig.~\figsref{fig:phase_diagrams}{(b), (d), (f)}.

\section{RG network}
Now that we have a base network capable of accurately determining the Chern number of disordered $4 \times 4$ lattices, the next step is generalizing to larger lattices.
We do this by training an ``RG" network that transforms an $8 \times 8$ lattice to a $4 \times 4$ one.
For this network, we use another 9-layer convolutional ResNet, whereby each layer consists of several ``kernels"---small blocks which are convolved with the lattice, incorporating periodic boundary conditions into the network structure (see Appendix~\ref{sec:rg_structure} for more details on the network's architecture).
This type of neural network is ideal for an RG-like operation because it operates on neighboring sites and transforms them into an effective site, regardless of the system size.
Each layer is also reminiscent of a renormalization group in that it is spatially local.

For the training of this RG network, we generated $10^7$ disorder realizations of $8 \times 8$ lattices, and calculated the Chern number for each of them using the Bott index method.
The RG network takes these samples as input, keeping the same input structure of $8 \times 8 \times 3$ parameters per sample, and outputs an effective $4 \times 4$ lattice: a list of $4 \times 4 \times 3$ parameters.
This effective reduced lattice is then combined with the  previously trained $4 \times 4$ base network to form a compound network, illustrated in a thick outline in Fig.~\figref{fig:block_diagram}{a}, which, as a whole, takes a disordered $8 \times 8$ lattice and returns an estimate of its Chern number. The resulting Chern number is compared to the correct one, and the network is trained to minimize the difference between the two, under the constraint that the base network's weights are not allowed to change.

This process alone can be used to train an RG network to high accuracy, 98.72\%, when evaluating $8 \times 8$ lattices.
The crucial question, though, is what happens when trying to apply this network to a larger lattice.
The simplest test is taking a $16 \times 16$ lattice, applying the RG network to obtain a $8 \times 8$ lattice, and then applying it again to obtain a $4 \times 4$ lattice.
We find that  the Chern number accuracy of this repeated network drops significantly (to 72.49\%).
This behavior is not surprising: the RG network has over $9 \times 10^6$ free parameters, and its training has no generalization requirement, so it has a very low probability of converging to a generalizable solution. 

To fix this issue, we initialize the RG network, prior to its training, to perform a na\"ive decimation mapping: constructing the smaller output lattice by simply averaging adjacent sites in each channel of the larger input.
Initializing the network in this way can be thought of as using the decimation mapping as a zeroth-order approximation of our desired function to be performed by the RG network (details on the initialization procedure are given in Appendix~\ref{initialization}).
The main step in our training procedure, which is training on the actual Chern numbers of $8 \times 8$ lattices, is then a refinement of this approximation.
As long as the network is only allowed to make small changes beyond na\"ive decimation, we expect it to stay generalizable.

\begin{figure}
    \centering
    \includegraphics[width=\linewidth]{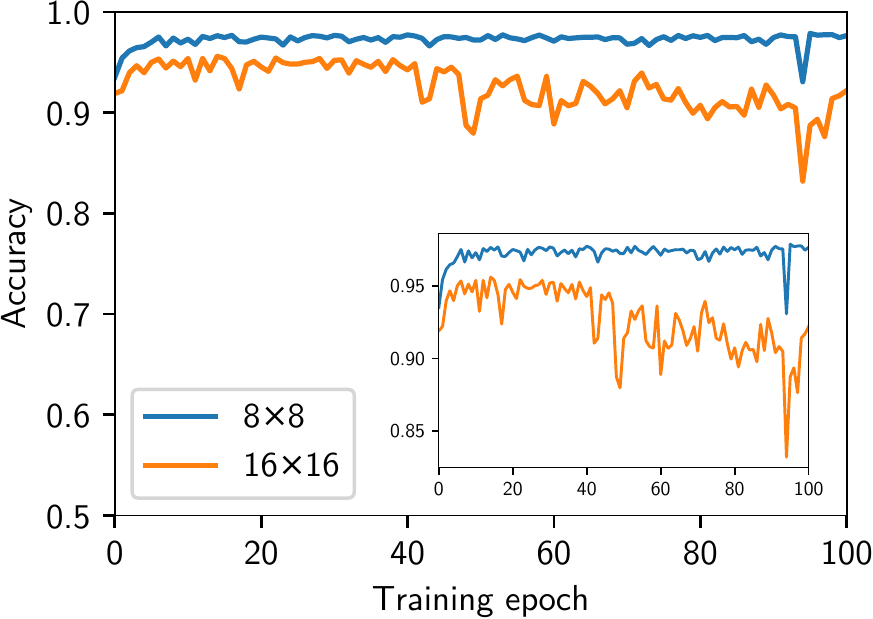}
    \caption{Accuracy of the RG network on $8 \times 8$ and (by applying the network twice in succession) $16 \times 16$ lattices, as a function of how long the network has been training. The accuracy for $16 \times 16$ initially increases, reaches a maximum at epoch 14, then declines.}
    \label{fig:RG}
\end{figure}

The training process for the RG network is plotted in Fig.~\ref{fig:RG}.
After each epoch---exposure to all of the samples in the $8 \times 8$ training set---the weights were saved and the current network was tested against a testing set of $8 \times 8$ lattices (blue curve).
In addition, the double-RG layer was tested against a testing set of $16 \times 16$ lattices (orange curve). Thus, the two curves in the figure represent the performance of the same RG network on multiple scales.
The initial accuracies---93.52\% and 91.91\% for $8 \times 8$ and $16 \times 16$, respectively---are the accuracies of the initial decimation function.
The network's performance on $8 \times 8$ samples increases roughly monotonically, as expected since this is the target function.
The $16 \times 16$ performance, however, reaches a maximum and then declines.
Our interpretation of this result is that the RG network begins in a generalizable subspace, but as the training process seeks an optimal $8 \times 8$ accuracy, it eventually drifts out of this subspace.

Since our goal is an RG network that generalizes well to large scales, we halt the training process when the $16 \times 16$ accuracy is at its peak.
The resulting RG network has an accuracy of 97.64\% on $8 \times 8$ lattices---slightly less than the best accuracy we attained, due to the fact that the training process was halted.
The accuracy on $16 \times 16$ lattices is 95.60\%, a significant improvement over the uninitialized case. We emphasize that this performance is despite the system never being exposed to a $16 \times 16$ lattice during training.

\section{Conclusion}
We have demonstrated that neural networks can accurately compute the Chern numbers of disordered Hamiltonians on lattices of a fixed size, and that RG-like networks can resize lattices while preserving the Chern number.
In combination, these two types of networks can, in principle, evaluate the topological invariants of lattices of arbitrary size, including sizes that are not computationally tractable with previous methods.
Indeed, our proof-of-concept RG network generalizes well to lattices a factor of two larger than those on which it was trained.

Given this promising result, the natural question is how large the input lattice can become before the compound network's accuracy decays beyond the point of usefulness. Unfortunately, this question is logistically difficult to answer. As discussed in Appendix~\ref{trainResults}, an appropriate disorder scaling convention must be found to counteract disorder averaging at larger scales. However, finding a viable convention is computationally expensive, as it requires slowly generating many $32 \times 32$ or larger lattices and computing their Chern numbers via the Bott index. Further study is needed to determine a more efficient means of solving this step; otherwise, our method's use remains restricted to systems for which a valid disorder scaling is known. For this reason, we constrained our scope to the $16 \times 16$ demonstration, as it is the simplest result that conveys the 
viability of our approach.

However, the accuracy does drop between the $16 \times 16$ and $8 \times 8$ scales, and we have no guarantee that it does not drop further for larger lattices.
A possible way to address this, which is left to future study, is an additional training stage on $16 \times 16$ samples.
We envision using two copies of the RG network in sequence, constrained to be identical throughout the training process, which reduce the input size by an overall factor of 4. This training step should further refine the RG block's ability to generalize to larger system sizes.

Beyond 2D systems and physics in general, iterated resizing networks represent an under-explored tool in computer science with three significant benefits. One is their ability, as we have demonstrated, to be applicable over many input sizes.
The second is the ability to reach high accuracies on inputs outside of the training distribution, which may be helpful in cases where training samples are more computationally difficult to produce (as is the case for our $16 \times 16$ lattices) or for which training data cannot be easily collected. Finally, our modular design offers a degree of interpretability. Most neural networks are ``black boxes" with very limited means of determining why they act the way they do. Our network with $8 \times 8$ inputs, for instance, can be examined at the intermediate $4 \times 4$ scale to study the network's interpretation of ``RG" (see Appendix~\ref{interpretability} for details). These benefits and extensions to our concept suggest that the use of base networks and iterated resizing blocks could prove useful in and beyond computational physics.

Models and related functions used in this research were written in Python using the machine learning packages Keras~\cite{Chollet2015Keras} and TensorFlow~\cite{Tensorflow2015}. Samples were generated using MATLAB. Our full code is available at \url{https://github.com/giladmargalit/ML-topology}.

\section*{Acknowledgment}
We are grateful to Oren Pereg, Moshe Wasserblat, Roman Beliy, and Itsik Adiv for helpful discussions.
The work at Weizmann was supported by the European Union's Horizon 2020 research and innovation programme (Grant Agreement LEGOTOP No. 788715), the DFG (CRC/Transregio 183, EI 519/7-1), ISF Quantum Science and Technology (2074/19), the BSF and NSF (2018643).
TPB acknowledges financial support from NSERC and FRQNT.

\appendix

\section{Structure and Training: Base Network}
\label{sec:base}

The base network is the final part of our network which receives a small ($4\times 4$ lattice points) input and returns the Chern number as an output.  Here we describe the details of this network.

The base network is a 9-layer convolutional ResNet~\cite{He2016ResNet} with skip connections between layers 2 and 3, 4 and 5, and so on. Each 2D convolutional layer has 256 kernels of size $3 \times 3$, with ReLU activation~\cite{Nair2010ReLU} and zero-padding (periodic padding is only used for the RG network). The output from these layers (of size $4 \times 4 \times 256$) is then combined via a max-pooling layer, flattened, and then passed through a dense layer with a single node and linear activation. This outputs the predicted Chern number.

The network was trained on an NVIDIA GeForce RTX 2080Ti GPU using $10^8$ samples of $4 \times 4 \times 3$ disordered lattices labeled with their corresponding Chern numbers (calculated via the Bott method~\cite{loring_disordered_2011,yi-fu_coupling-matrix_2013}) in mini-batches of size 640. We used stochastic gradient descent (SGD) with a momentum of 0.9 and an initial learning rate of 0.01. No learning rate decay was used, but a callback function reduced the learning rate by a factor of 0.1 each time the validation loss did not reach a new minimum for 50 consecutive epochs. When the learning rate would have been reduced to $10^{-7}$, the training instead terminated.

\subsection{Sample Distribution}
\label{samples}

Each $4 \times 4 \times 3$ sample lattice is produced in two steps. First, in each of the 3 channels (chemical potential $\mu$, hopping anisotropy $t^-$, and pairing energy $\Delta$), a random disorder value is generated for each site, chosen uniformly from a disorder range specific to each parameter type and subtracting the average so that each channel's disorder averages to 0. The disorder ranges used in this work are $[-0.15,0.15]$ for $\mu$, $[-0.1,0.1]$ for $t^-$, and $[-0.025,0.025]$ for $\Delta$. These disorder ranges were chosen to allow a noticeable impact on Chern number while not resulting in phase diagrams that are vastly different from the clean diagram.

Next, we choose an average value for each of the 3 channels and add it to every site in the lattice. Unlike the disorder distribution, these average values are not selected uniformly in each range. In order to use training time more efficiently, our training set is boundary-biased, meaning that the distribution of average site values is biased toward points close to the phase boundaries, where the Chern number is most strongly affected by disorder (only the training set, not the testing set, is biased in this way; the latter is generated uniformly within the average and disorder ranges). Specifically, the training set distribution of samples is
\begin{equation}
\label{eq:dist}
D(\mu,t^-,\Delta) = \frac{N}{\alpha + x(\mu,t^-)},
\end{equation}
where $N$ is a normalization factor, $\alpha=0.03$ is a constant that indicates the level of bias, and $x(\mu,t^-)$ is the distance between the point $(\frac{\mu}{2},t^-)$ and the nearest clean phase boundary, (the vertical and diagonal lines where the Chern number changes in Fig.~2a in the main text; we set $t^+$ equal to 1). Note that since the phase boundaries of clean systems do not depend on $\Delta$, the distribution of average $\Delta$ values is uniform. $\mu$, $t^-$ and $\Delta$ are constrained to the ranges $[-4,4]$, $[-2,2]$, and $[0.1,0.2]$, respectively, and are subjected to the boundary-biasing distribution within those ranges.

\begin{figure}
     \includegraphics[width=\linewidth]{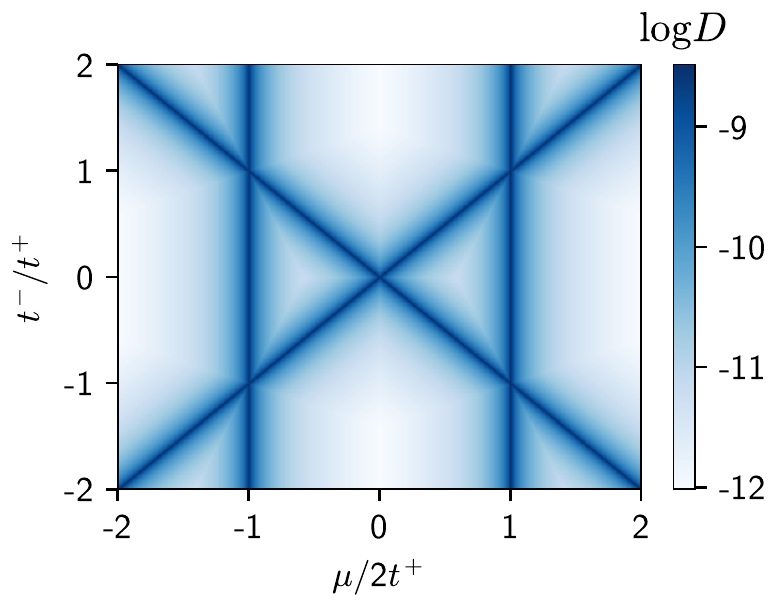}
\caption{Biased distribution of training samples according to their distance from the phase boundaries (see Eq.~\eqref{eq:dist}), with  $\alpha=0.03$.} \label{fig:samples_distribution}
\end{figure}

\subsection{Custom Loss}
\label{customLoss}

We make use of a custom loss function in all of our training throughout this research. The function is
\begin{equation}
\label{eq:loss}
L(y_{\text{pred}} - y_{\text{true}}) = \abs{y_{\text{pred}} - y_{\text{true}}} +(y_{\text{pred}} - y_{\text{true}})^2,
\end{equation}
where $y_{\text{pred}}$ and $y_{\text{true}}$ are the predicted value and true value of the label, respectively. For the base network, we apply this loss on each channel independently, then add the results to form an overall loss function.

This loss has the advantage of harshly punishing large deviations from the labels, as a mean square loss does, while also being unforgiving of small errors in the latter stages of training, since the absolute value function is larger for values close to zero. We found substantially-improved accuracy on early tests when using this loss function compared with mean-square or linear loss functions, as well as with discrete losses such as categorical cross-entropy. In later tests with our final training set and disorder distribution, the difference between this loss and conventional losses was less pronounced, but the loss was not modified further.

\section{Structure and Training: RG Network}
\label{sec:rg_structure}

The structure of the RG network is nearly identical to the base network: a 9-layer ResNet. All parameters are the same as the base network except as indicated below.

The input layer is of size $8 \times 8 \times 3$, and the output is of size $4 \times 4 \times 3$. In order to reduce the dimensions of the lattice, the first convolutional layer has its stride value set to $(2,2)$, meaning that the kernels only evaluate on sites with odd numbers as both of their indices. The sites with even index still contribute to the output, however, since each is within several $3 \times 3$ kernels centered at nearby odd-indexed sites. All layers after the first use the standard $(1,1)$ stride, and in addition, the number of kernels in each layer after the first is 512 so as to keep the total number of parameters consistent between rows.

Rather than use zero padding, we employ periodic padding (described in more detail in Sec.~\ref{periodic}) on the 9 main convolutional layers. After the 9 main layers there is a single convolutional layer with only 3 kernels of size 1, which results in an output of the desired shape; the padding on this layer is irrelevant since the kernel cannot extend beyond the lattice boundary, so it is set to zero padding for simplicity.

\begin{figure}
     \includegraphics[width=\linewidth]{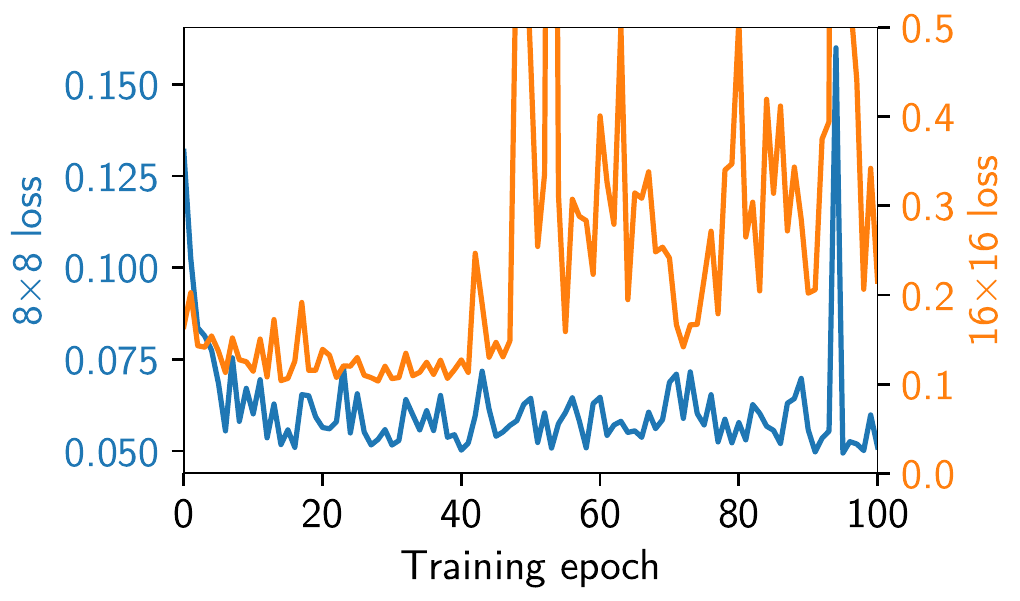}
\caption{Training loss for $8 \times 8$ (blue) and $16 \times 16$ (orange) evaluations of the RG network.} \label{fig:loss}
\end{figure}

For further illustration of the training process, the loss is plotted in Fig.~\ref{fig:loss} for the same training run whose accuracy is depicted in Fig.~\ref{fig:RG}.

\subsection{Initialization}
\label{initialization}

As described in the main text, before the RG network can be trained to classify Chern numbers, it must first be initialized such that it approximates a decimation mapping. This preliminary step is motivated by the fact that decimation---equivalent to an average pooling layer of pool size $(2,2)$---performs somewhat well on both $8 \times 8$ and $16 \times 16$ lattices (with accuracy 93.52\% and 91.91\%, respectively). We hypothesized that an RG network initialized in this way would retain some of the generalizability (functionality at multiple lattice scales) of the decimation function in the early stages of its training to preserve the Chern number when it changes the system size.

The goal of the initialization process is to begin with a network that performs a simple decimation, meaning that it merely averages nearby parameters, creating a coarse-grained lattice.  Practically, it is easier to train the network to perform this decimation rather than setting its weights by hand.  We therefore begin with a normal distribution of kernel weights and biases and the RG ResNet is trained on $5 \cdot 10^7$ samples of $8 \times 8$ lattices. Rather than being labeled by their Chern number, these samples' labels are the $4 \times 4$ lattices that would result from a decimation mapping being applied to them. The loss during training is the custom loss described in Sec.~\ref{customLoss} applied to each channel and site of the output, comparing them to the corresponding channel and site of the label, and then summing the result with all sites and channels weighted equally. The training procedure is identical to that of the base network, except that the learning rate is set to 0.0005 and not adjusted during training.

The samples used in the initialization stage form a much wider distribution than those used in the Chern number training. Specifically, each site and channel of a sample contains a uniformly-chosen random number in the range $[-5,5]$. It is critical that this domain includes, but is much larger than, the parameter range of the base and RG network's Chern number training.  If the initialization step is trained only on lattices from the same distribution as the Chern-learning step, the initialized network does not approximate a general decimation, but rather a decimation that has been overfitted to function only on $8 \times 8$ lattices in our specific input domain. Naturally, this prevents it from generalizing successfully to $16 \times 16$ lattices.

\subsection{Chern Training and Results}
\label{trainResults}

In the main training step, a compound network, composed of the initialized RG network feeding into the base network, is trained on $5 \cdot 10^7$ samples of $8 \times 8$ lattices and their Chern numbers. The base network's weights are frozen, so that they do not update and only the RG network is optimized. The training process is identical to the base network case except for a learning rate of 0.0005, which is not changed during training.

The samples are distributed identically to the $4 \times 4$ samples described in Sec.~\ref{samples}, except that the range of disorder in $\mu$ in both training and testing samples is increased from $[-0.15,0.15]$ to $[-0.3,0.3]$. This counteracts disorder averaging, which causes the effect of disorder on the Chern number to become less significant at larger lattice scales. Without scaling the disorder in this way, a trivial RG mapping that ignores disorder altogether can still achieve high accuracy. Since we are free to define how lattice parameters scale in our renormalization scheme, we choose to increase this disorder value in order to prevent the RG network from converging to this trivial map. This ensures that our compound network will be able to solve for the Chern number for large lattices in the strong-disorder regime.

The compound network is saved after each epoch. The RG network corresponding to each epoch is then loaded and tested on an $8 \times 8$ testing set of size $10^6$. Additionally, each network is placed in a compound network with another copy of itself and a base network, allowing it to evaluate $16 \times 16$ lattices. This requires only that the input layer of each RG network be removed, that the two be placed in sequence, and that a new input layer of size $16 \times 16 \times 3$ be added. The latter compound network is tested on a $16 \times 16$ testing set of size $10^6$. The accuracy of these tests as a function of epoch number are given in Fig.~3 in the main text and in Fig.~\ref{fig:periodicCompare} below.

\begin{figure}
     \includegraphics[width=\linewidth]{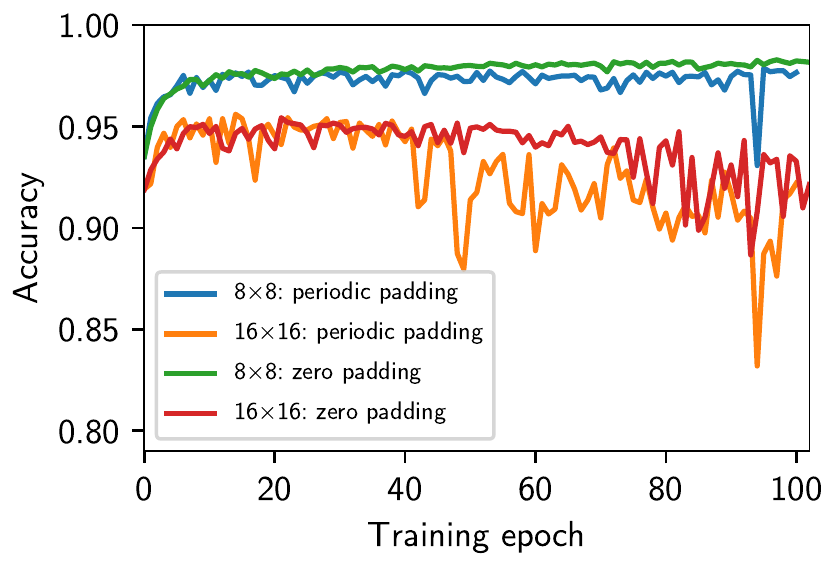}
\caption{Accuracy as a function of epochs trained for $8 \times 8$ and $16 \times 16$ evaluations of the RG network.} \label{fig:periodicCompare}
\end{figure}

\subsection{Periodic Padding}
\label{periodic}

Unlike in the image processing applications that are the usual use case for convolutional neural networks, the lattices we use as inputs represent periodic systems. As such, it is physically reasonable to have the network interpret the lattice in such a way that periodicity is preserved.

To do this, we define a periodic padding. When the $3 \times 3$ kernel is centered at a site on the edge or corner of the input lattice, some of the positions in the kernel will be outside the lattice. Instead of interpreting the values of these nonexistent so-called pad sites as uniformly zero in each channel, as most convolutional networks do, we treat their value as being what it would be if the lattice was periodic; that is, equal to the corresponding site on the other side of our input lattice. Practically, this is accomplished by inserting a Lambda layer---a layer that performs a predefined function---before each convolution, which adds a periodic pad. The input to each convolutional layer is thus 2 sites larger in each dimension, corresponding to the 1-site-thick pad. The padding procedure of the convolutional layers is set to ``valid", meaning that the border sites are not allowed to be the center sites of kernels, which reduces the output of the convolutional layer by 2, back to its correct size.

The effect of periodic padding on the computed Chern number accuracy of $8 \times 8$ and $16 \times 16$ lattices is shown in Fig.~\ref{fig:periodicCompare}. For unknown reasons, periodic padding causes the accuracy to grow more slowly in the $8 \times 8$ case and to fall more rapidly in the $16 \times 16$ case for later epochs of training. However, in our testing, it provided the best single saved network at evaluating $16 \times 16$ lattices (95.60\% accuracy, compared to 95.43\% accuracy for the best non-periodic network). Periodic padding also aids in interpretability, since it removes a notable and unphysical distortion along the system boundary when using the RG network as a resizing mapping (see Sec.~\ref{interpretability}). The frequent layer resizing causes RG training time to increase by about 40\%, however.

\section{Properties of the RG Network}
\label{interpretability}

Our network has a two-stage structure: a down-sizing RG block is followed by a Chern number calculation block. This structure allows us to examine the properties of the RG network alone, without looking at the Chern number itself. In this section, we describe some basic observations on the nature of this network.

\subsection{Points Mapping}

\begin{figure*}[t!]
     \includegraphics[width=\linewidth]{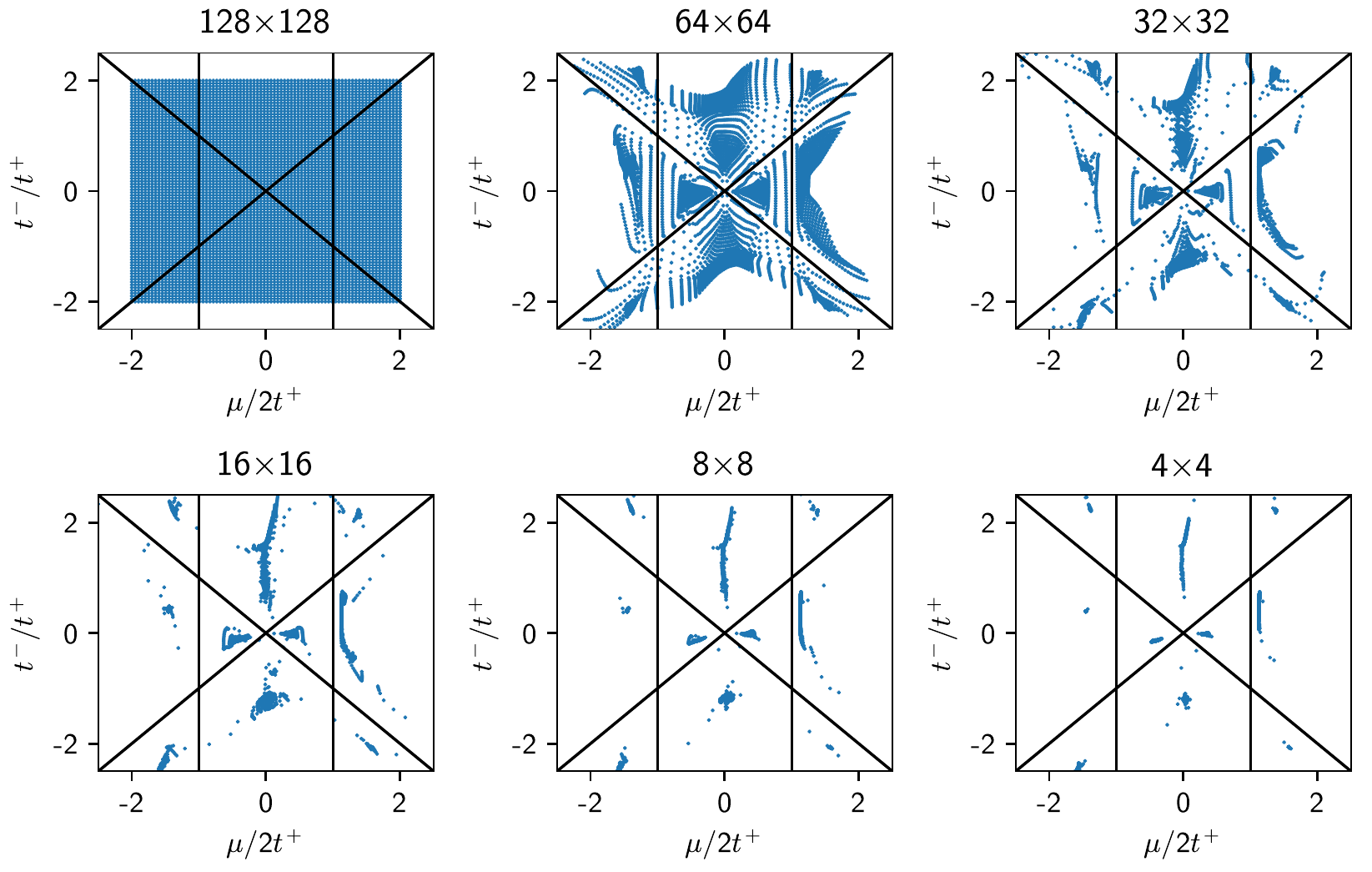}
\caption{Visualization of the RG mapping. Each point in the $128 \times 128$ plot corresponds to a single clean lattice, characterized by its chemical potential $\mu$ and the hopping anisotropy $t^{-}$. The solid black lines represent the phase boundaries in the clean case. As the RG network is applied iteratively, the points cluster in each phase, demonstrating that the RG mapping actively preserves the Chern number.} \label{fig:points_mapping}
\end{figure*}

As a first step, we ask how the RG transformation acts in parameter space. We know that the Chern number is mostly preserved, but that alone does not mean the lattice does not drift far away in parameter space---many points have the same Chern number despite being distant from each other in parameter space. However, our training starts from simple decimation, which leaves parameter space invariant, so we expect the output lattice's parameters not to stray too far from those of the input lattice.

To check this notion of locality in parameter space, we generated clean $128\times 128$ lattices with values of $\mu$ and $t^{-}$ evenly spaced in the interval $[-2,2]$ (as usual, we set $t^{+}=1$). Simple decimation leaves these points exactly where they started in the 2D parameter space of $\mu,t^{-}$. Our network, however, acts slightly differently. The $64 \times 64$ plot in Figure.~\ref{fig:points_mapping} shows the distribution of points after applying the RG network. Each point in this plot corresponds to a clean lattice which resulted from the RG mapping being applied to a clean lattice from the $128\times 128$ plot. Rather than being evenly spaced, we observe a high density of points deep inside the phases, and generally a much lower density close to the phase transition lines. Our interpretation is that the network is trying to avoid regions with large uncertainty, where the error in the Chern number would be much higher. Physically, the topological gap in these regions is small, and therefore the Chern number becomes harder to determine reliably.

The network has the freedom to choose an asymmetric transformation, as we do not force any symmetry; hence, the point plots do not obey the mirror symmetries of the phase boundaries in general. Moreover, different runs of the training procedure may randomly lead to different asymmetries.

As the RG mapping is applied iteratively, the points cluster into fixed points. Since this mapping has been trained to preserve Chern number, we expect at least one fixed point per phase region, and this is indeed what we find. If this holds for most disorder realizations as well, it implies that in large-scale implementations of this method, the inputs will be clustered and effectively classified by the time they are reduced to the $4 \times 4$ scale. Thus, in these large-scale implementations, it is likely that the base network could be substituted with a much smaller, simpler network, further improving runtime.

\subsection{Mapping of a Single Impurity}

\begin{figure*}[t!]
     \includegraphics[width=0.45\linewidth]{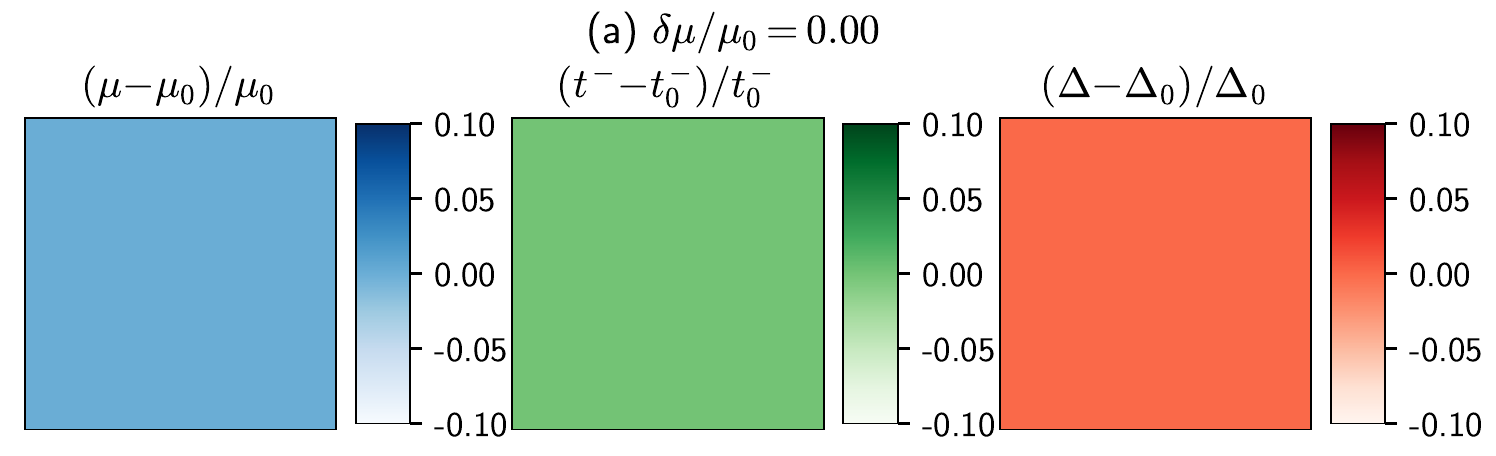}
     \includegraphics[width=0.45\linewidth]{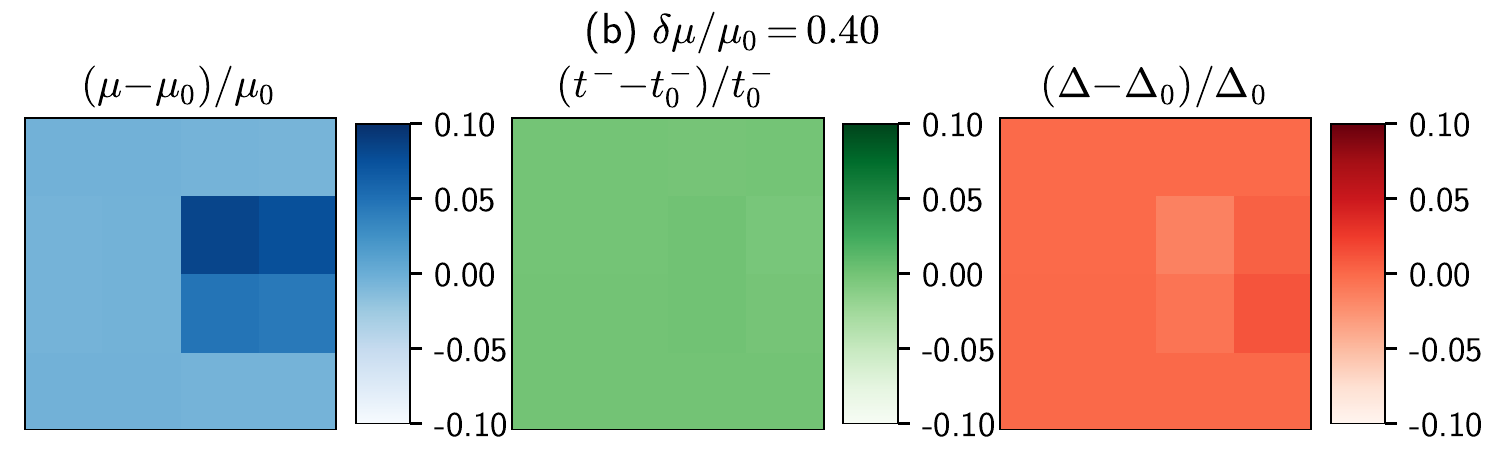}
     \includegraphics[width=0.45\linewidth]{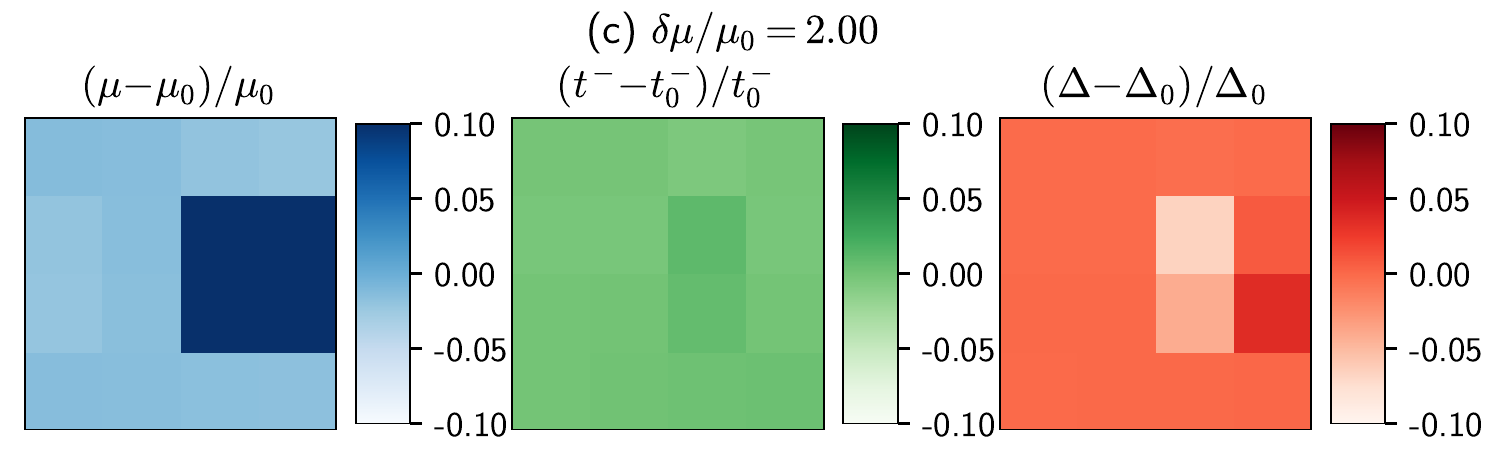}
     \includegraphics[width=0.45\linewidth]{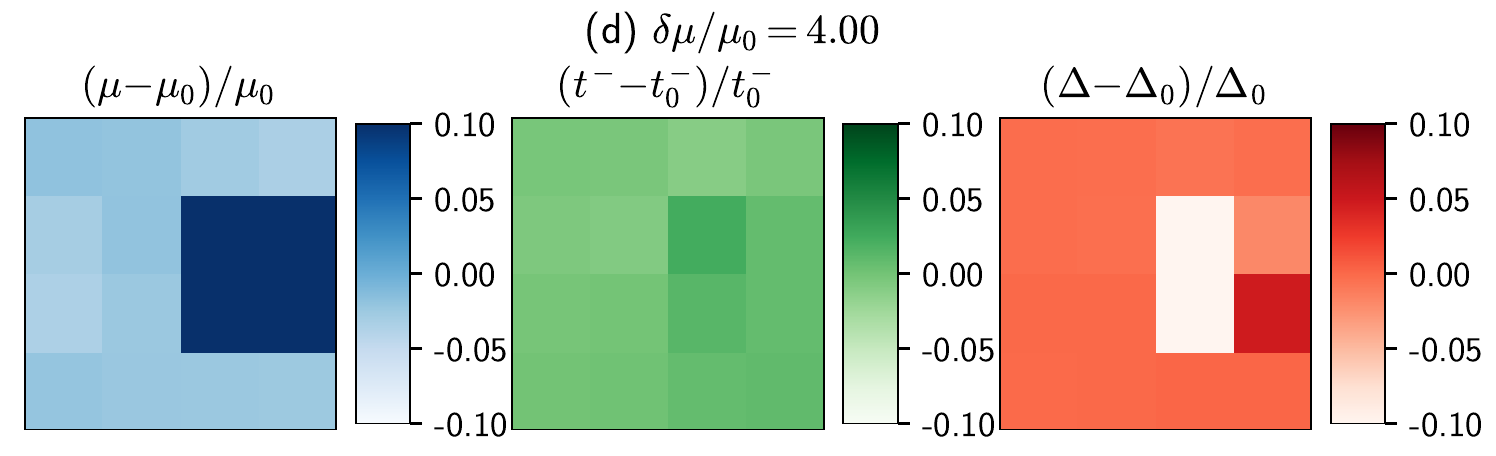}
\caption{Propagation of a single impurity in the chemical potential from an $8\times 8$ lattice down to the renormalized $4\times 4$ lattice calculated by the RG network. The values used are $\mu_0=0.5$, $t^{-}_0=1$, $\Delta_0=0.15$. The impurity $\delta\mu$ is placed at the site $(4, 6)$, and its values are indicated on the different panels.} \label{fig:mu_impurity}
\end{figure*}

Going one step further, we would like to understand how simple configurations are mapped under the RG network. As an example, we consider a single impurity in an otherwise clean $8\times 8$ lattice. The impurity is in the form of a different chemical potential, $\mu_0+\delta\mu$, at a specific site, where all the other sites have chemical potential $\mu_0$. Specifically, here we take $\mu_0=0.5t^{+}$ and place the impurity at the site $(4, 6)$. 

In Fig.~\ref{fig:mu_impurity} we show the $4\times 4$ lattices resulting from applying the RG network to this impurity configuration, for four values of the impurity $\delta\mu$. For each $\delta\mu$, we plot the values of $\mu,t^{-},\Delta$ at each point in the $4\times 4$ lattice, subtracting their initial values to get a difference map.

Without an impurity ($\delta\mu$=0), the $4\times 4$ lattice is clean, as its $8\times 8$ ancestor, and the parameters closely match the original ones---see Fig.~\figref{fig:mu_impurity}{a}. As $\delta\mu$ is increased, we observe the impurity propagating to the $4\times 4$ lattice in a \emph{local} way. First, the impurity mostly affects the sites close to the location of the original impurity (if one were to coarse-grain the $8\times 8$ lattice into a $4\times 4$ one). Second, the propagation is mostly local also in channel space: the variation in $\mu$ affects mostly the renormalized $\mu$, and its effect on the renormalized $t^{-},\Delta$ is weaker. This demonstrates that although the Chern number is a global property, the RG-like transformation that preserves it is mostly local in nature.

\bibliography{library}

\end{document}